# Interlayer band-to-band tunneling and negative differential resistance in van der Waals BP/InSe field effect transistors

*Quanshan Lv, Faguang Yan, Nobuya Mori, Wenkai Zhu, Ce Hu, Zakhar R. Kudrynskyi, Zakhar D. Kovalyuk, Amalia Patanè[*] and Kaiyou Wang[*]*

Mr. Q. Lv, Dr. F. Yan, Mr. W. Zhu, Mr. C. Hu, Prof. K. Wang

State Key Laboratory of Superlattices and Microstructures, Institute of Semiconductors, Chinese Academy of Sciences, Beijing 100083, China

College of Materials Science and Opto-Electronic Technology, University of Chinese Academy of Science, Beijing 100049, P. R. China

E-mail: kywang@semi.ac.cn.

Dr. Z. Kudrynskyi, Prof. A. Patanè

School of Physics and Astronomy, University of Nottingham, Nottingham NG7 2RD, UK

E-mail: amalia.patane@nottingham.ac.uk

Prof. N. Mori

Division of Electrical, Electronic and Information Engineering, Graduate School of Engineering, Osaka University, Japan

Prof. Z. Kovalyuk,

Institute for Problems of Materials Science, The National Academy of Sciences of Ukraine, Chernivtsi Branch, Chernivtsi 58001, Ukraine

Prof. K. Wang

Beijing Academy of Quantum Information Sciences, Beijing 100193, China

**Abstract**

Atomically thin layers of van der Waals (vdW) crystals offer an ideal material platform to realize tunnel field effect transistors (TFETs) that exploit the tunneling of charge carriers across the forbidden gap of a vdW heterojunction. This type of device requires a precise energy band alignment of the different layers of the junction to optimize the tunnel current. Amongst two-dimensional (2D) vdW materials, black phosphorus (BP) and indium selenide (InSe) have a Brillouin zone-centered conduction and valence bands, and a type II band offset, both ideally suited for band-to-band tunneling. Here, we demonstrate TFETs based on BP/InSe heterojunctions with diverse electrical transport characteristics: forward rectifying, Zener-tunneling and backward rectifying characteristics are realized in BP/InSe junctions with different thickness of the BP layer or by electrostatic gating of the junction. Electrostatic gating yields a large on/off current ratio of up to $10^8$ and negative differential resistance at low applied voltages ($V \sim 0.2$V). These findings illustrate versatile functionalities of TFETs based on BP and InSe, offering opportunities for applications of these 2D materials beyond the device architectures reported in the current literature.

# 1. Introduction

The progressive miniaturization of electronic devices has propelled several technologies to higher performance and efficiency, but further progress and innovative solutions to global challenges require a shift from traditional approaches towards transformative material systems and integration technologies.[1,2] Atomically thin layers of van der Waals (vdW) crystals and their heterostructures,[3-5] generally referred to as two-dimensional (2D) materials, offer opportunities to study and exploit quantum phenomena for a wide range of applications.[6-10] These crystals have strong covalent atomic bonding in the 2D planes and weak vdW interaction between the layers, which enable the fabrication of stable thin films down to the atomic monolayer thickness and stack them into multi-layered heterostructures.[11-13] The science of these 2D systems is developing rapidly with important technological breakthroughs emerging from recent studies.

Amongst the extended family of 2D systems, the metal chalcogenide indium selenide, InSe,[14-18] and the elemental compound black phosphorous, BP,[19-24] have received increasing attention. These two semiconductors have electronic properties distinct from those of other 2D materials, such as transition metal dichalcogenides (TMDCs), including a higher electron mobility beneficial for field effect transistors (FETs).[15,17,25] Also, the direct (*e.g.* BP) and "quasi-direct" (*e.g.* InSe) band gap energy of the 2D layers increases markedly with decreasing layer thickness down to a single layer, leading to an optical spectrum ranging from the violet to the infrared wavelength range.[26,27] Thin layers of InSe and BP has been used in a diverse range of functional devices, ranging from quantum point contacts,[18] giant quantum Hall effect devices,[16,28,29] and high polarization-sensitive photodetectors.[23,30] However, the electronic properties of the InSe/BP heterostructure[31-34] remain still largely unexplored. Of particular interest is the opportunity to exploit tunneling across this heterojunction as BP and InSe present a number of attractive features: they both have a Brillouin zone-centered conduction band (CB) and exhibit a type II band offset,[31-33] which are well suited to control and exploit the transmission of charge carriers between different bands of the heterostructure for the realization of a tunnel-FET (TFET). This type of device concept has been demonstrated using junctions based on graphene and other 2D materials,[7,35] including $WSe_2/SnSe_2$ heterojunctions[36] and tunnel diodes based on $MoS_2/WSe_2$ with a symmetric dual-gate architecture.[37] However, for many 2D materials, such as TMDCs, the band edges are located at the K-point of the Brillouin zone. Thus, a precise alignment of the different layers of the junction is required to optimize the tunnel current across the heterostructure.

Here, we report on tunnel-FETs (TFETs) based on a BP/InSe heterostructure contacted with graphene electrodes and capped with a hBN layer acting as an effective encapsulating layer and dielectric for electrostatic gating. Compared to TFETs based on lateral structures,[35,38] in our TFETs the graphene, BP and InSe layers are vertically stacked leading to thin tunnel barriers for the charge carriers. This is beneficial for controlling and modulating the tunnel current across the junction. We demonstrate the operation of these TFETs at low applied voltages (< 0.5 V) and a diverse range of electrical characteristics controlled by the field effect, including negative differential resistance (NDR) due to interlayer band-to-band tunneling (BTBT). In particular, we show that with increasing the BP layer thickness from ~ 6 nm to ~ 60 nm, the heterojunction changes its properties revealing a transition from a forward rectifying *pn*-diode behaviour to Zener tunneling and backward rectification. The transition between these diverse functionalities can also be achieved in a single device through a dual gate modulation approach.

## 2. Results and discussion
### 2.1. BP/InSe heterojunction

**Figure 1**a shows a schematic and an optical image of a typical BP/InSe heterojunction onto a SiO$_2$/Si substrate (thickness $t_{SiO2}$= 300 nm). The junction is capped with hBN and contacted with graphene electrodes. The Si-substrate serves as a bottom gate (*bG*) of the FET. Ta/Au contact pads on two graphene electrodes serve as source (*s*) and drain (*d*). The InSe-layer is grounded (*s*) and the drain voltage, $V_{ds}$, is applied to the BP (*d*). The top hBN capping layer (thickness $t_{hBN}$ = 20 nm) is covered with graphene, serving as a top gate electrode (*tG*), and is contacted with Ta/Au. In this device, the graphene sheets serve as charge extraction electrodes. In particular, the bottom graphene electrode acts to screen dopant states from the SiO$_2$ substrate, thus providing a clean interface.

Figure 1b shows the energy band alignment out of equilibrium of bulk BP and InSe. This is derived from the values of the electron affinity ($\chi_{BP}$ = - 4.4 eV, $\chi_{InSe}$ = -4.6 eV) and band gap energy of the bulk crystals at room temperature ($E_{BP}$ = 0.3 eV, $E_{InSe}$ = 1.26 eV),[17,39] resulting into a type-II heterojunction with a small offset of 0.10 eV between the CB of InSe and the valence band (VB) of BP. We note that the neutrality point of the Dirac cone of graphene ($\chi_g$ = -4.5 eV) is close to the VB maximum of BP and to the CB minimum of InSe, thus facilitating the injection and extraction of holes or electrons from the graphene electrodes into/from the BP/InSe heterostructure under appropriate applied source-drain voltages and/or electrostatic gating. Due to the unintentional doping of the crystals during the growth, the BP and InSe bulk crystals are *p*-type and *n*-type doped, respectively (Supplementary Information S3). For InSe, the electron density is $n = 10^{15}$ cm$^{-3}$ at room temperature. For BP, the carrier concentration tends to increase from $p = 3.9 \times 10^{15}$ cm$^{-3}$ to $p = 2.7 \times 10^{17}$ cm$^{-3}$ with increasing the layer thickness from $t_{BP}$ <10 nm to $t_{BP}$ >50 nm. A similar p-type doping of BP and dependence of the hole density on the BP layer thickness were reported before.[40,41]

**2.2. Diverse electrical transport characteristics for BP/InSe heterojunctions**

We have fabricated and investigated devices with different thickness of the BP layer ($t_{BP}$ <10 nm, 20-30nm and > 50 nm) and InSe flakes of similar thickness $t_{InSe}$ = 15-20 nm (Supporting Information S5 and S7). Figure 1c-e shows the dependence of the current, $I_{ds}$, on the source-drain voltage, $V_{ds}$, at room temperature ($T$ = 300K) without any applied back- or top-gate voltage (*e.g.* $V_{bG}$ =0 V and $V_{tG}$ =0 V). As shown in Figure 1c, for $t_{BP}$ ~ 6 nm, the device shows forward rectifying characteristics, as expected for a *p*-BP/*n*-InSe junction. When the BP layer thickness is increased to $t_{BP}$ = 30nm (Figure 1d), both the forward and reverse current increase. In particular, the large increase of the current in reverse bias indicates a breakdown of the junction and Zener tunneling of electrons from the VB of *p*-type BP into the CB of *n*-type InSe. A further increase of the BP layer thickness to $t_{BP}$ ~ 60 nm results into backward rectifying characteristics (Figure 1e) with an exponential growth of the current in reverse bias suggesting dominant Zener tunneling. The dependence of the rectification ratio on the thickness of the BP layer is shown in the Supporting Information Figure S1. The three different transport characteristics shown in Figure 1 indicate that the behaviour of the BP/InSe junction is very sensitive to the thickness of the BP layer. We note that quantum confinement of carriers in BP can be neglected for all the flake thicknesses ($t_{BP}$ > 5 nm) considered in our structures. On the other hand, the work function of BP depends on the layer thickness $t_{BP}$: with increasing $t_{BP}$ from 6 nm to 60 nm, it was found that the Fermi level comes closer to the valence band edge; in particular, BP can become a degenerate *p*-type semiconductor for thick layers.[39] This change, which remains still largely unexplored, can be exploited to tune the energy band alignment and transport characteristics when BP is combined with other 2D materials.[40,41] Since electrostatic gating represents an effective way to control and modulate the carrier density, we now examine the properties of the BP/InSe junction under different applied gate voltages.

### 2.3. Electrostatic gating of BP/InSe heterojunctions

The three transport characteristics shown in Figure 1 for different BP/InSe junctions can be realized by electrostatic gating in a single device. Here we focus on a junction with $t_{BP}$ = 30 nm and $t_{InSe}$ = 20 nm. **Figure 2**a-f shows the dependence of the measured current intensity, $|I_{ds}|$, on the source-drain voltage, $V_{ds}$, at room temperature for different applied back-gate voltages, $V_{bG}$, and top-gate voltages, $V_{tG}$. These dependencies are also illustrated in the colour scale plots of $|I_{ds}|$ versus $V_{ds}$ and $V_{bG}$ in Figure 2g-i: different behaviours of the junction are observed by our dual gating approach. The transfer characteristics at fixed $V_{ds}$ and their temperature dependence are in the Supporting Information S4.

Figures 2a-c show the $|I_{ds}|$-$V_{ds}$ curves for $V_{bG}$ increasing from -50 to 0V by increments of 5V, while keeping $V_{tG}$ constant at -8V (Figure 2a), 0V (Figure 2b) and +8V (Figure 2c). For $V_{tG}$ = -8V (Figure 2a) and $V_{tG}$ = 0V (Figure 2b), the $|I_{ds}|$-$V_{ds}$ curves show the characteristic forward-type rectifying behaviour of a *pn*-junction. The rectification ratio increases by one order of magnitude with decreasing the back gate voltage from $V_{bG}$ = 0 to -50. For $V_{tG}$ = +8V (Figure 2c), decreasing $V_{bG}$ from 0 to -50V induces a transition from a backward- to a forward-type rectifying behaviour, which can be explained

by the bipolar behavior of BP[20,42]: The positive top gate voltage $V_{tG}$ modulates the unintentionally p-type doped BP to n-type, resulting into an n-BP/n-InSe junction. A negative back gate voltage can counterbalance this effect. Thus, p-BP/n-InSe ($V_{bG}$ < -20V) and n-BP/n-InSe ($V_{bG}$ > -20V) junctions can be both realized at $V_{tG}$ = +8V.

Figures 2d-e show the $|I_{ds}|$-$V_{ds}$ curves for $V_{bG}$ increasing from 0V to +50V by increments of 5V, while keeping $V_{tG}$ constant at -8V (Figure 2c), 0V (Figure 2d) and +8V (Figure 2e). Under these conditions, the current $|I_{ds}|$ increases with increasing $V_{bG}$ from 0V to +50V. For $V_{tG}$ = -8V (Figure 2c), the $|I_{ds}|$-$V_{ds}$ curves exhibit backward-type rectifying characteristics; also, the current exhibits a steep rise at $V_{ds}$ < 0 V, suggesting a breakdown of the junction and Zener tunneling of electrons from p-BP to n-InSe. For $V_{tG}$ = 0V (Figure 2e) and $V_{tG}$ = +8V (Figure 2f), the positive top gate and back gate voltages modulate p-type BP to n-type, resulting into an n-BP/n-InSe junction. Thus, a backward-type rectifying $|I_{ds}|$-$V_{ds}$ curves and a larger forward current are observed at $V_{tG}$ = 0V compared to $V_{tG}$ = -8V. The dependence of $|I_{ds}|$ on $V_{ds}$ and $V_{bG}$ at different $V_{tG}$ is shown in the colour scale plots of Figure 2g-h. In summary, diverse transport characteristics can be realized by dual gating. Also, similar dependences of $|I_{ds}|$ on $V_{ds}$ and $V_{bG}$ are observed in devices with different thickness of the BP layer although for different values of $V_{bG}$ and/or $V_{tG}$ (Supporting Information S5 and S7).

To examine further the tunnel current across the BP/InSe junction, we zoom in the $|I_{ds}|$-$V_{ds}$ curves in the low bias $V_{ds}$-region. **Figure 3** shows the $|I_{ds}|$-$V_{ds}$ curves at $V_{bG}$ = -50V and $V_{tG}$ = 0V for different temperatures $T$ from 10 to 300K. A well-defined peak in $|I_{ds}|$-$V_{ds}$ and corresponding region of NDR are clearly seen in forward bias at low temperatures ($T$ < 60K). The NDR is observed at low voltage $V_{ds}$ ~ 0.2V and is followed at high voltages by a steep increase of the current. In the region of NDR the current density is of up to $100^2$ A/m$^2$. Such a well-defined peak in $|I_{ds}|$-$V_{ds}$ and high current density were never reported in heterojunctions based on BP or TMDCs.[35, 39] We assign the region of NDR to Zener tunneling across the $p$-BP/$n$-InSe junction. When the temperature increases to $T = 60$K, the NDR tends to broaden revealing a switching on/off behavior of the tunnel current. Following a further increase of temperature, this switching behavior disappears and the NDR becomes weaker and broader, disappearing when the temperature increases above $T = 150$K. The measured temperature dependence reveals two competing conduction mechanisms due to tunneling and thermionic emission. At low temperatures ($T < 60$K), the contribution of thermionic emission is small and the current is dominated by tunneling, leading to NDR. We assign the on/off switching behavior of the current to thermally assisted ionization of charge carriers onto localized states at the heterojunction interface. With increasing temperature the NDR region broadens due the thermal broadening of the carrier distributions in the layers and scattering of charge carriers by phonons. Also, the increasing contribution of thermionic emission tends to mask the NDR region in $|I_{ds}|$-$V_{ds}$.

The tunnel current and region of NDR in $I_{ds}$-$V_{ds}$ are strongly dependent on electrostatic gating. **Figure 4**a shows the dependence of the NDR on the negative back gate voltage, $V_{bG}$, at $T = 50$K. With increasing the back gate voltage from $V_{bG} = $ -50V to 0V, the NDR tends to shift to higher $V_{ds}$ and its peak-to-valley ratio decreases. In particular, for $V_{bG} < $ - 15V, the current is strongly suppressed and no NDR is observed. This behavior is illustrated in the colour scale plot of the differential conductance $G_{ds}$ versus $V_{bG}$ and $V_{ds}$ for $V_{tG} = 0$V in Figure 4b and can be understood by considering the energy band diagrams of Figure 4c-d. Here we examine a BP/InSe heterojunction under a forward (c) and reverse (d) applied voltage. In both cases, the $p$-BP layer is highly doped with a Fermi level lying below the VB edge, corresponding either to a large negative back gate and/or a negative top gate voltage. For a small forward source-drain voltage applied to the junction, electrons tunnel from the CB of *n*-type InSe into the empty states of the VB of *p*-type BP (Figure 4c-left panel). When $V_{ds}$ is further increased, the energy window for tunneling decreases and correspondingly the current decreases. A further increase of bias leads to a monotonic increase of the current due to diffusion of majority carriers across the junction (Figure 4c-right panel). A different behaviour can be instead expected in reverse bias (Figure 4d). Under a small negative bias, the Fermi level of BP is within the forbidden gap of InSe and tunneling across the junction cannot occur (Figure 4d-left panel). When the reverse bias voltage is increased further (Figure 4d-right panel), an energy window is created for electrons to tunnel from the filled VB states of BP into the empty CB states in InSe, leading to an increase in the current. At low temperatures, we measured an on/off ratio that exceeds $10^8$. This is 1 to 2 orders of magnitude higher than that of TFETs based on TMDCs.[35,41]

## 3. Conclusions

In conclusion, we have demonstrated band-to-band tunneling in BP/InSe heterostructures by using a dual-gate device architecture. The electrical transport properties of the devices vary from forward rectification to Zener-like tunneling and backward rectification with increasing the thickness of the BP layer. These diverse device functionalities can also be realized in a single device by electrostatic gating. The type II band alignment between BP and InSe enables the realization of a Zener diode with a large on/off ratio of the current, which reaches a value of up to $10^8$ under low applied biases. Notably, a robust NDR region can be observed under a forward bias at low temperatures and different gate voltages, confirming that the measured electrical current is mainly due to band-to band tunneling across the interface of the BP/InSe heterojunction. This work demonstrates versatile functionalities of tunnel-FETs based on BP and InSe, opening opportunities for further research and applications such as polarization-sensitive photodetectors and impact avalanche diodes. Also, different designs based on the incorporation of a tunnel barrier (*e.g*. hBN) at the BP/InSe interface could be envisaged to minimize carrier thermal diffusion and enhance the region of NDR for potential applications in microwave electronics.

## 4. Experimental Section

The dry transfer method for stacking the layers inside a glove box is as follows. The few-layer flakes of InSe and BP were mechanically exfoliated using adhesive tape from bulk crystals. The InSe was grown by the Bridgman method; the bulk BP crystal was purchased from Hefei Kejing Materials Technology. The tape with the InSe or BP flakes was adhered to a polymethyl methacrylate (PMMA) stamp on a glass slide to facilitate handling and identification of specific flakes by optical microscopy. The target InSe sheet was then transferred onto the top of the first graphene electrode (drain) on a $SiO_2$/Si substrate. Using the same method, the BP sheet was transferred on top of the InSe sheet to create a BP/InSe/graphene stack. Hence, another graphene layer was stamped onto the BP sheet to form the top electrode (source). The graphene/BP/InSe/graphene stack was capped with a 20 nm-thick hBN layer and a graphene layer was then dry transferred onto the hBN for electrostatic gating. The thicknesses of all flakes were measured by atomic force microscopy (AFM).

Graphene was grown by chemical vapor deposition (CVD) on a copper foil substrate and then transferred onto the $SiO_2$/Si substrate to form microstamps. The graphene layers were shaped into electrodes using electron-beam lithography, oxygen plasma etching and buffered oxide etching (BOE).

Au/Ta electrodes (50/10 nm) were fabricated on the SiO$_2$/Si substrate prior to the transfer using standard photoetching, thermal evaporation and lift-off.

The $I_{ds}$ - $V_{ds}$ curves were measured by an Agilent Technology B1500A Semiconductor Device Analyzer. The measurement resolution of the Semiconductor Device Analyzer was down to 0.1 fA and 25 µV.

**Supporting Information**

Additional data on the samples and their properties are in the Supplementary Information.

**Acknowledgements**

This work was supported by National Key R&D Program of China 2017YFA0303400, by the NSFC (Grant No. 61774144), by Chinese Academy of Sciences (Grant No. QYZDY-SSW-JSC020, XDPB12, and XDB28000000), the Engineering and Physical Sciences Research Council [grant number EP/M012700/1]; the European Union's Horizon 2020 research and innovation programme Graphene Flagship Core 2 under grant agreement number 785219; The University of Nottingham; The National Academy of Sciences of Ukraine; The Leverhulme Trust [RF-2017-224].

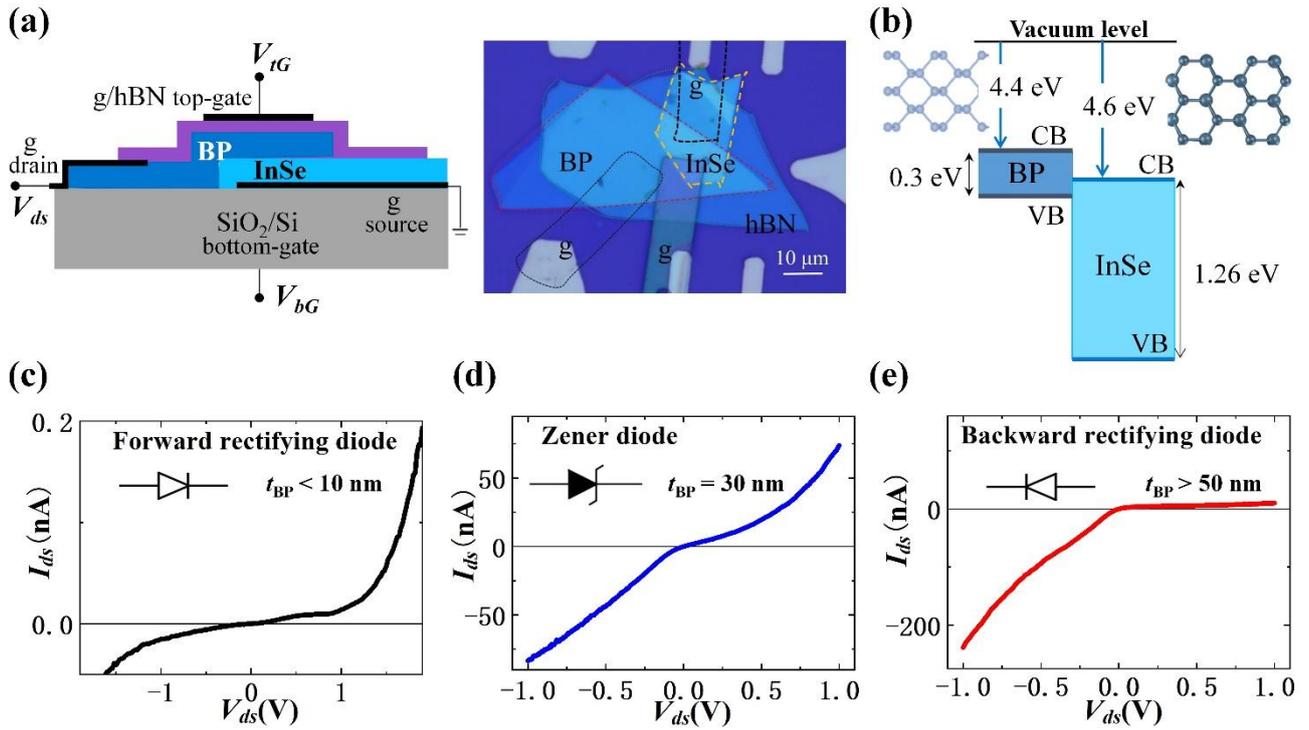

**Figure 1.** Tunnel field effect transistors (TFETs) based on BP and InSe

(a) Schematic and optical image of a typical BP/InSe heterojunction device ($t_{BP}$ = 30 nm and $t_{InSe}$ = 20 nm). (b) Energy band alignment out of equilibrium of bulk BP and InSe. (c-e) Diverse diode properties of BP/InSe heterojunctions based on BP with different layer thickness ($t_{BP}$ < 10, 30 and > 50 nm) at room temperature ($T$ = 300K) without any applied back-gate and top-gate voltages (*e.g.* $V_{bG}$ = 0V and $V_{tG}$ = 0V).

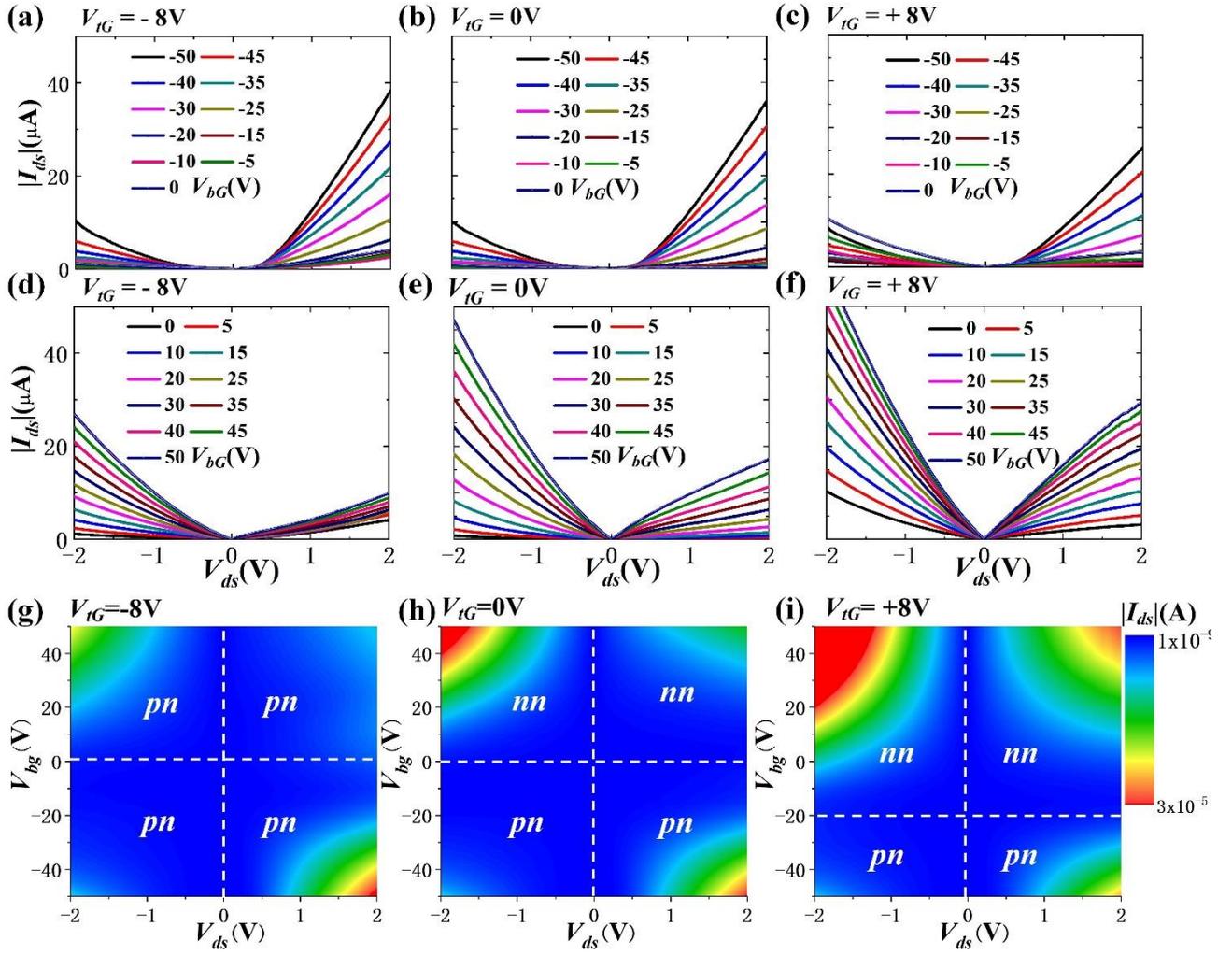

**Figure 2.** Electrical transport characteristics of BP/InSe tunnel field effect transistors

Current-voltage $|I_{ds}|$-$V_{ds}$ curves at room temperature for a BP/InSe heterojunction with $t_{BP}$ = 30 nm and $t_{InSe}$ = 20 nm. (a-c) Different $|I_{ds}|$-$V_{ds}$ curves in each panel correspond to a bottom gate voltage $V_{bG}$ that varies from 0 to -50 V in steps of 5 V, while keeping the top gate voltage, $V_{tG}$, constant at -8V (a), 0V (b) and +8V (c). (d-f) Different $|I_{ds}|$ -$V_{ds}$ curves in each panel correspond to a bottom gate voltage $V_{bG}$ that varies from 0 to +50 V in steps of 5 V, while keeping the top gate voltage, $V_{tG}$, constant at -8V (d), 0V (e) and 8V (f). (g-i) Colour scale plot of $|I_{ds}|$ versus $V_{bG}$ and $V_{ds}$ for $V_{tG}$ = -8V (g), 0V (h) and 8V (i). The colour scale ranges from $10^{-9}$ A (blue) to $3\times10^{-5}$ A (red).

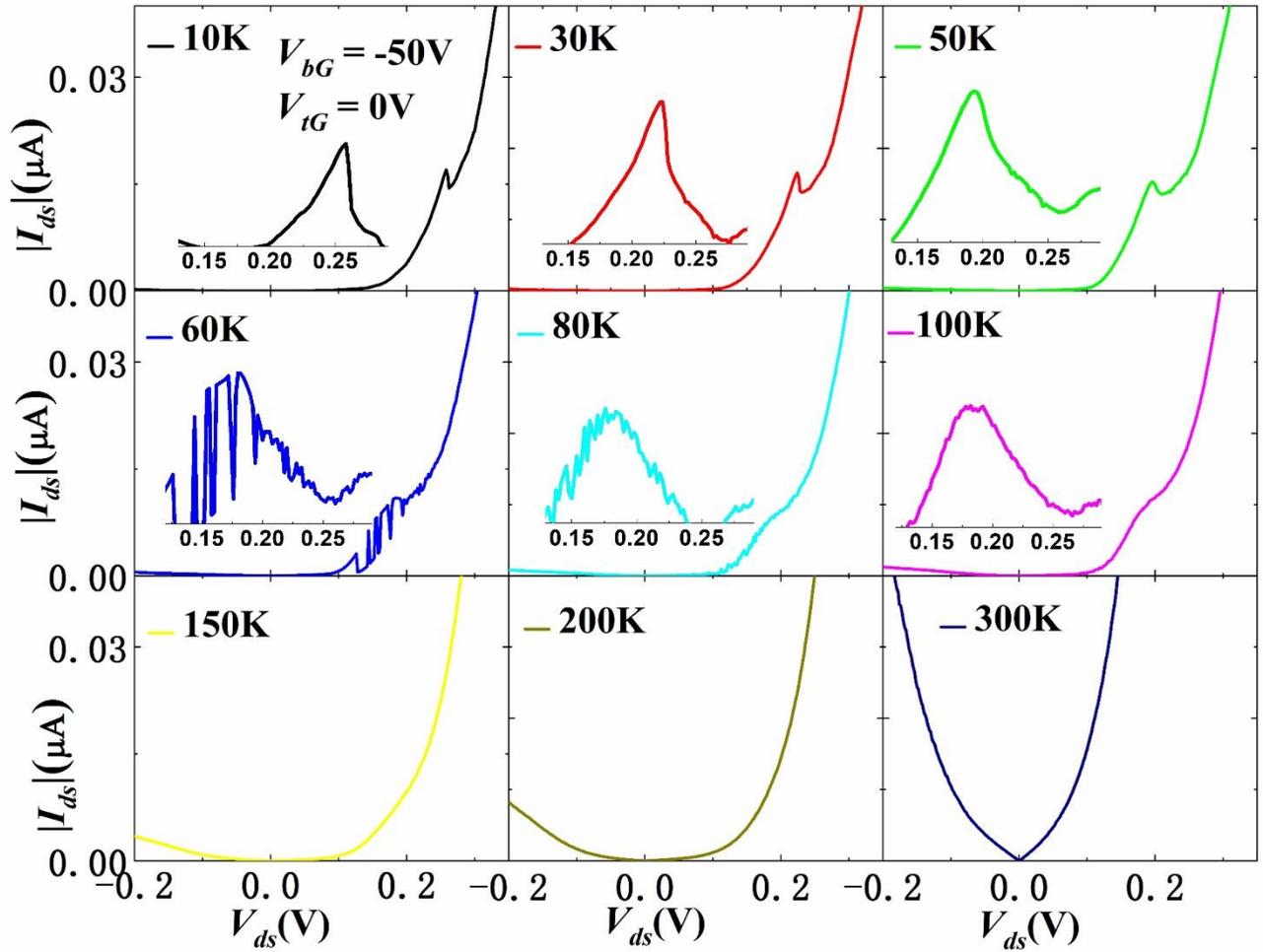

**Figure 3.** Negative differential resistance in BP/InSe tunnel field effect transistors

Current-voltage $|I_{ds}|$-$V_{ds}$ curves for a BP/InSe heterojunction device with $t_{BP}$ = 30 nm and $t_{InSe}$ = 20 nm. The $|I_{ds}|$-$V_{ds}$ are shown for a bottom-gate voltage $V_{bG}$ =-50V and a top-gate voltage $V_{tG}$ =0 V at various temperatures from $T$ = 10K to 300K. Insets: region of negative differential resistance in $|I_{ds}|$-$V_{ds}$ after subtraction of the background current.

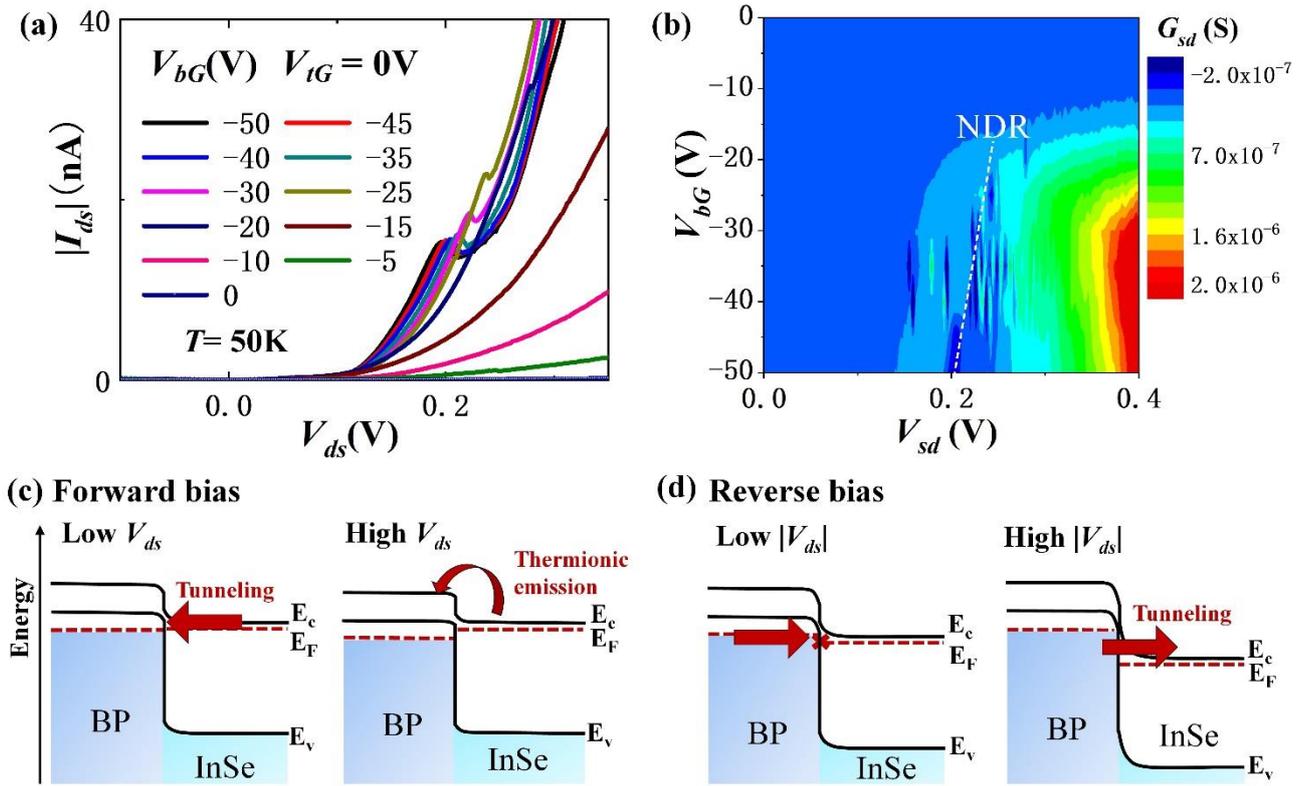

**Figure 4.** Electrostatic gating and negative differential resistance

(a) Current-voltage $|I_{ds}|$-$V_{ds}$ curves for a BP/InSe heterojunction device with $t_{BP}$ = 30 nm and $t_{InSe}$ = 20 nm. The $|I_{ds}|$-$V_{ds}$ curves are shown for different bottom-gate voltages, $V_{bG}$, at a top-gate voltage $V_{tG}$ = 0 V and temperature $T$ = 50K. $V_{bG}$ varies from 0 to -50 V in steps of 5 V. (b) Colour scale plot of the differential conductance $G_{ds}$ versus $V_{bG}$ and $V_{ds}$ for $V_{tG}$ = 0V. The colour scale ranges from -2×10⁻⁷ S (blue) to 2×10⁻⁶ S (red). (c-d) Energy band diagrams under a forward (c) and reverse (d) bias.

# Supplementary information
# Interlayer band-to-band tunneling and
# negative differential resistance
# in van der Waals BP/InSe field effect transistors

*Quanshan Lv, Faguang Yan, Nobuya Mori, Wenkai Zhu, Ce Hu, Zakhar R. Kudrynskyi, Zakhar D. Kovalyuk, Amalia Patanè[*] and Kaiyou Wang[*]*

Mr. Q. Lv, Dr. F. Yan, Mr. W. Zhu, Mr. C. Hu, Prof. K. Wang

State Key Laboratory of Superlattices and Microstructures, Institute of Semiconductors, Chinese Academy of Sciences, Beijing 100083, China

College of Materials Science and Opto-Electronic Technology, University of Chinese Academy of Science, Beijing 100049, P. R. China
E-mail: kywang@semi.ac.cn.

Dr. Z. Kudrynskyi, Prof. A. Patanè

School of Physics and Astronomy, University of Nottingham, Nottingham NG7 2RD, UK

E-mail: amalia.patane@nottingham.ac.uk

Prof. N. Mori

Division of Electrical, Electronic and Information Engineering, Graduate School of Engineering, Osaka University, Japan

Prof. Z. Kovalyuk,

Institute for Problems of Materials Science, The National Academy of Sciences of Ukraine, Chernivtsi Branch, Chernivtsi 58001, Ukraine

Prof. K. Wang

Beijing Academy of Quantum Information Sciences, Beijing 100193, China

## S1. Dependence of the rectification ratio on the thickness of BP

We calculated the rectification ratio for devices with different thickness of the BP layer. With increasing thickness from 6 nm to 60 nm, the rectification ratio decreases from 2 to 0.04, indicating a transition from a forward- to a backward-type rectifying behaviour.

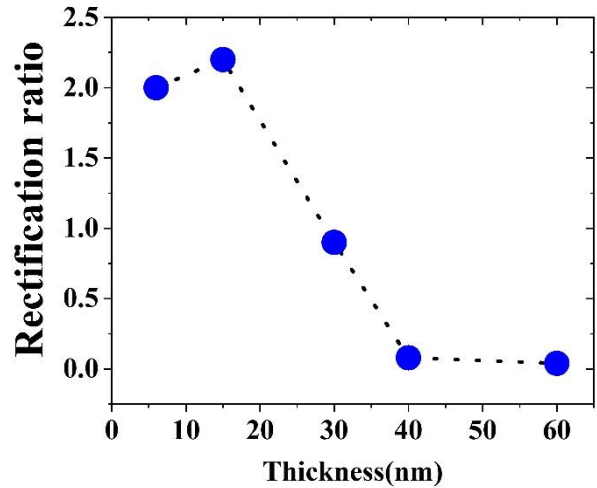

**Figure S1**. Rectification ratio for BP/InSe FETs on a SiO$_2$/Si substrate with different thickness of the BP layer ($T$ = 300K).

## S2. Back gate modulation of BP and InSe layers

We measured the room temperature transfer characteristics of FETs where individual BP and InSe layers are sandwiched between two graphene electrodes. For the BP-based FET, with sweeping the back gate voltage from -60V to 60V, the current first decreases, reaching a minimum at a positive gate voltage $V_{bG}$ ~ 10 V, and then increases, demonstrating a bipolar behavior for BP (Figure S2a). For the InSe-based FET, the transfer curve of Figure S2b shows that the current increases with increasing gate voltage, indicating a *n*-type conductivity for InSe. The current on/off ratio is smaller than for BP.

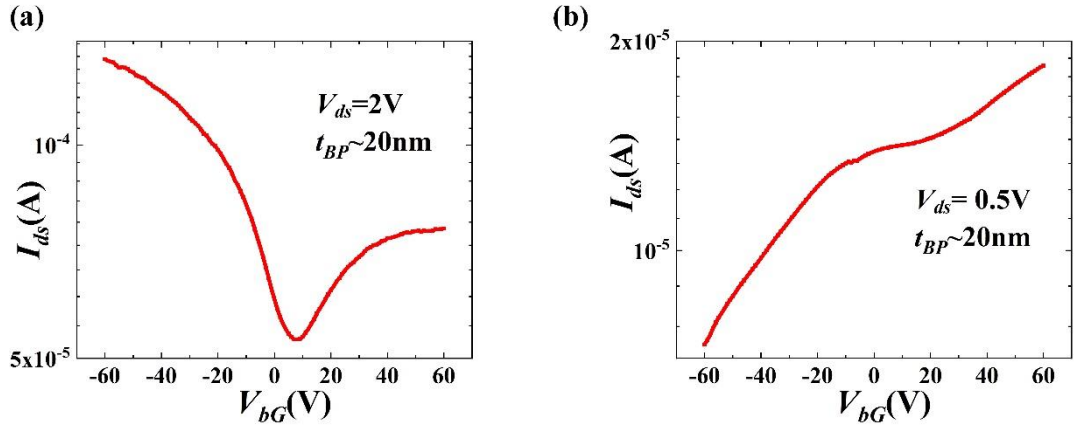

**Figure S2**. (a) Transfer curve of a graphene/BP/graphene FET on a SiO$_2$/Si substrate with $t_{BP}$ = 20 nm. The dependence is shown at room temperatures for a source-drain voltage $V_{ds}$ = 2V. (b) Transfer curve of a graphene/InSe/graphene FET on a SiO$_2$/Si substrate with $t_{InSe}$ = 20 nm. The dependence is shown at room temperatures for a source-drain voltage $V_{ds}$ = 0.5V.

## S3. Carrier density and mobility of bulk BP and InSe

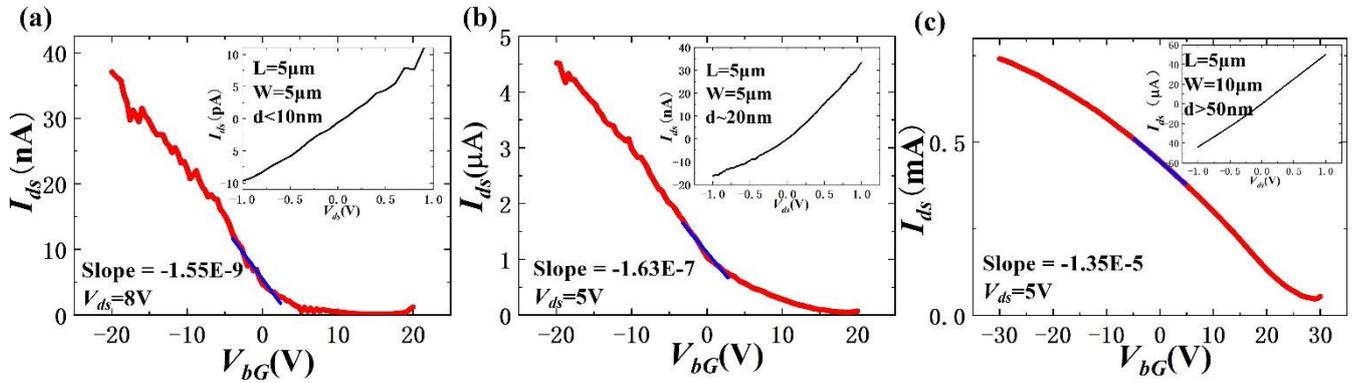

**Figure S3**. Transfer curves of a BP/hBN FET on a SiO$_2$/Si substrate with different thickness of the BP at $T$ = 300K: $t_{BP}$ <10nm(a), ~20nm(b) and >50 nm(c). Insets: $I_{ds}$ - $V_{ds}$ curves of the BP/hBN FET with $V_{bG}$ = 0V.

The mobilities and densities of carriers in BP with three different thickness were estimated from measurements of the source-drain current, $I_{ds}$, versus the applied gate voltage, $V_{bG}$, in BP-based FETs on a Si/SiO$_2$ substrate (Figure S3). The estimated carrier mobility and hole density in these three BP flakes are $\mu = 0.016$ cm$^2$/Vs, $p = 3.9 \times 10^{15}$ cm$^{-3}$ ($t_{BP} < 10$ nm); $\mu = 2.73$ cm$^2$/Vs, $p = 4 \times 10^{16}$ cm$^{-3}$ (~ 20nm) and $\mu = 113.2$ cm$^2$/Vs, $p = 2.7 \times 10^{17}$ cm$^{-3}$ (> 50nm), respectively ($V_{bG} = 0$V and $T = 300$K).

The majority carriers in nominally undoped InSe are electrons. From Hall measurements, we derive an electron density $n = 10^{15}$ cm$^{-3}$ and electron mobility $\mu = 1000$ cm$^2$/Vs at $T = 300$K.

## S4. Transfer characteristics of the BP/InSe heterojunction

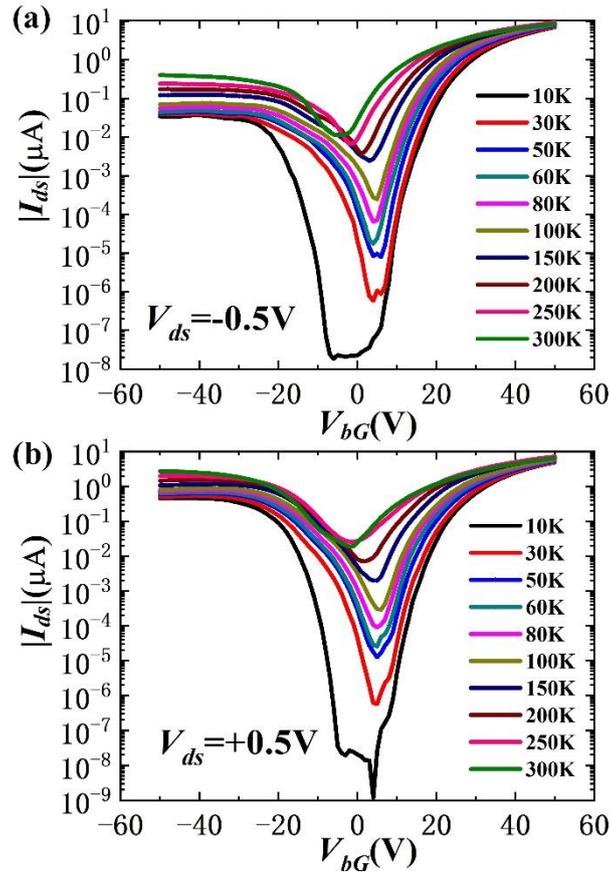

**Figure S4.** Modulation of the source-drain current by electrostatic gating. Dependence of the amplitude of the source-drain current, $|I_{ds}|$, on the back-gate voltage, $V_{bG}$, for a BP/InSe heterojunction

device with $t_{BP}$ = 30 nm and $t_{InSe}$ = 20 nm. The dependence is shown at various temperatures from $T$ = 10K to 300K for source-drain voltages $V_{ds}$ = -0.5V (a) and $V_{ds}$ = 0.5V (b).

## S5. BP/InSe heterostructure with a thin BP layer ($t_{BP}$<10 nm)

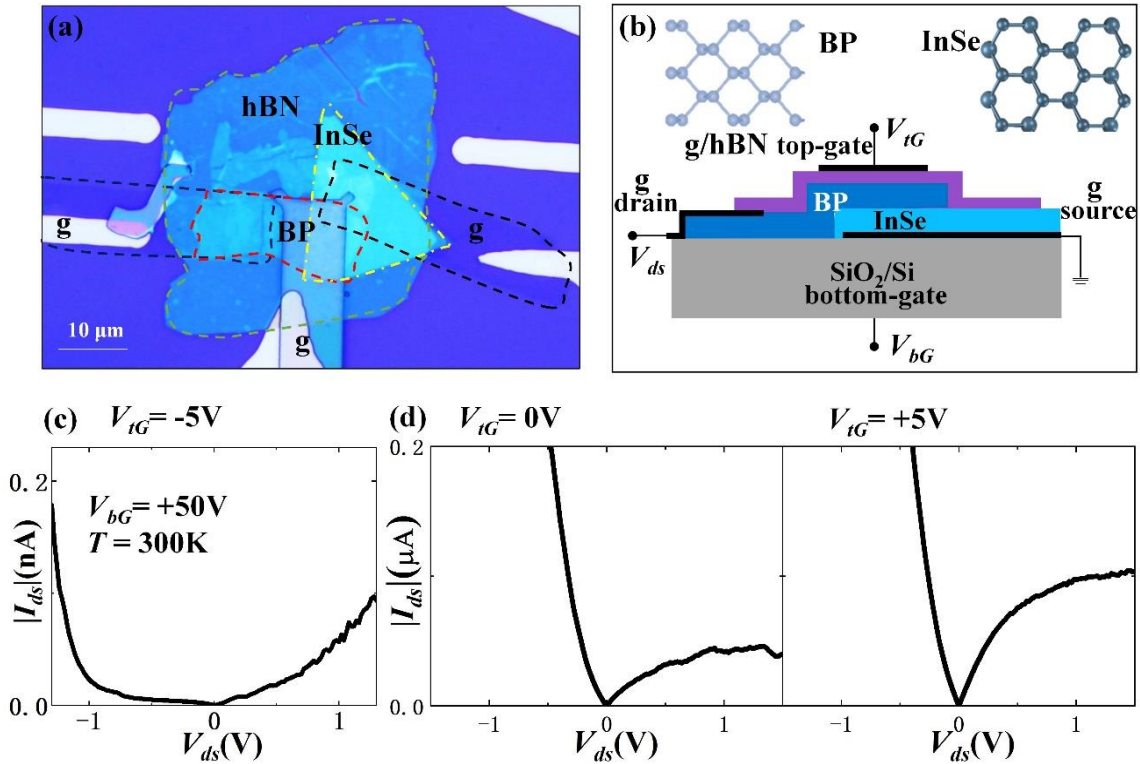

**Figure S5.** BP/InSe van der Waals heterostructure with a thin ($t_{BP}$<10 nm) BP layer. Optical image (a) and schematic layout (b) of a BP/InSe FET on a SiO$_2$/Si substrate with graphene electrodes and a hBN encapsulating layer. The inset in part (b) shows the in-plane crystal lattice of BP and InSe. (c-e) Current-voltage $|I_{ds}|$-$V_{ds}$ curves at top-gate voltages $V_{tG}$ = -5V (c) and $V_{tG}$ = 0, +5V (d) at room temperature ($T$ = 300K). The bottom-gate voltage is $V_{bG}$ = +50V for all panels.

Figure S5a-b shows the optical image and the schematic of the BP/InSe heterostructure with thickness of the BP ($t_{BP}$<10 nm) and InSe ($t_{InSe}$ = 20 nm). The thickness of the encapsulating layer

hBN is 20 nm. Figure S5c-d shows the measured dependence of the amplitude of the current, $|I_{ds}|$, on the source-drain voltage, $V_{ds}$, at $T = 300K$ for different applied top-gate voltages, $V_{tG}$, and a bottom-gate voltage $V_{bG} = +50V$. A noticeable feature of these data is the different form of the $|I_{ds}|$-$V_{ds}$ curves for $V_{tG} = -5V$ (Figure S5c) and $V_{tG} = 0, +5V$ (Figure S5d). At $V_{tG} = -5V$ (Figure S5c), the $|I_{ds}|$-$V_{ds}$ curve shows the characteristic forward-type rectifying behaviour of a *pn*-junction, *i.e.* for low applied voltages the current under a forward bias, $V_{ds} > 0$, is higher than for $V_{ds} < 0$. The current exhibits a steep rise at $V_{ds} < -1V$, suggesting a breakdown of the junction and Zener tunnelling of electrons from *p*-BP to *n*-type InSe. At $V_{tG} = 0, +5V$ (Figure S5d), the $|I_{ds}|$-$V_{ds}$ curves exhibit backward-type rectifying characteristics and a saturation behaviour at $V_{ds} > +0.5V$.

We examined the modulation of the current by the top gate voltage in the device with thinner BP layer. The $V_{tG}$-dependence of current $|I_{ds}|$ at low ($T = 11$ K) and room temperature ($T = 300K$) was measured at various $V_{bG}$ and $V_{ds}$. As shown in Figure S6a, at $T = 300K$, $V_{ds} = -0.5V$ and $V_{bG} = 0V$, the amplitude of the current, $I_{ds}$, reaches a minimum at a negative top-gate voltage $V_{tG} \sim -3$ V. The minimum of $|I_{ds}|(V_{tG})$ shifts to a lower $V_{tG} \sim -5$ V at $V_{bG} = +50$ V and to a higher $V_{tG} \sim -1$ V at $V_{bG} = -50$ V. A similar behavior is observed for $V_{ds} = +0.5V$ (Figure S6b). The transfer characteristics change with lowering the temperature from $T = 300$ to $11K$: at $V_{bG} = 0V$ and $T = 11K$, the current goes to zero; furthermore, at low temperature the modulation of the current by the top-gate voltage becomes stronger for $V_{tG} > -3V$ at $V_{bG} = +50$ V and for $V_{tG} < +1$ V at $V_{bG} = -50$ V. The subthreshold swing is defined as $SS = [d\log(|I_{ds}|)/dV_{tG}]^{-1}$ in units of mV/decade. A *SS* of ~100 mV/dec was obtained.

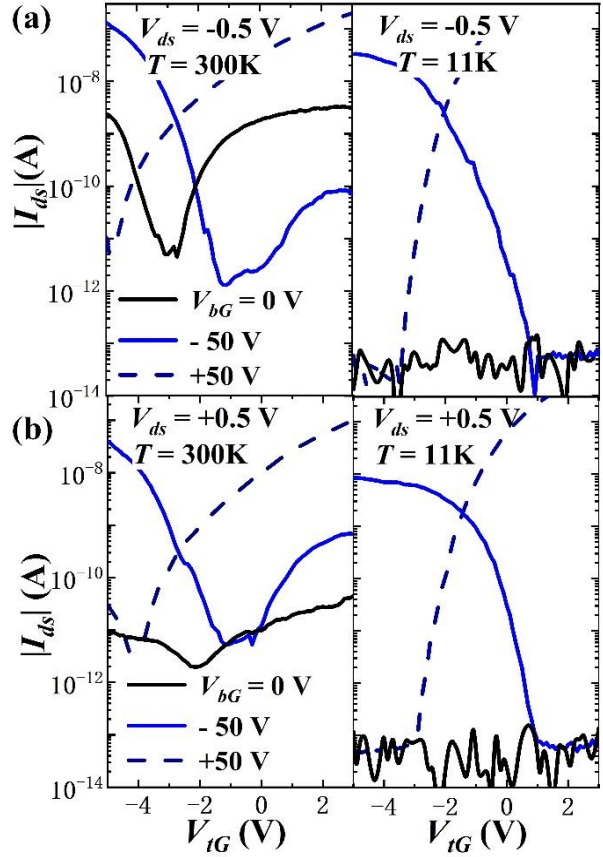

**Figure S6.** Modulation of the source-drain current by electrostatic gating. (a-b) Dependence of the amplitude of the source-drain current, $|I_{ds}|$, on the top-gate voltage, $V_{tG}$, at temperatures $T = 300K$ and $T = 11K$ for source-drain voltages $V_{ds} < 0$ (a) and $V_{ds} > 0$ (b). Different curves in each panel correspond to different bottom gate voltages $V_{bG}$.

## S7. BP/InSe van der Waals heterostructure with a thick ($t_{BP} > 50$ nm) BP layer.

Figure S7a-b shows the optical image and the schematic of the BP/InSe heterostructure with thickness of the BP ($t_{BP} > 50$ nm) and InSe ($t_{InSe} = 20$ nm). The thickness of the encapsulating layer hBN is 20 nm. Figure S7c-d show the $|I_{ds}|$-$V_{ds}$ curves as a function of varying back gate voltage ranging from 0 to 50V with an increment of 5V, while keeping the top gate voltage, $V_{tG}$, constant at -6V (Figure S7c), 0V (Figure S7d) and 6V (Figure S7e), respectively. For $V_{tG} = -8V$, the $|I_{ds}|$-$V_{ds}$ curves exhibit

backward-type rectifying characteristics and a saturation behaviour at $V_{ds} > 0$V; also, the current exhibits a steep rise at $V_{ds} < 0$ V, suggesting a breakdown of the junction and Zener tunnelling of electrons from $p$-BP to $n$-type InSe. For $V_{tG} = 8$ and 0 V, with the increases of back gate voltages, the $|I_{ds}|$-$V_{ds}$ curves exhibit symmetrical characteristics, which corresponding to a $n$-BP/$n$-InSe junction. This dependence is similar to that observed in BP/InSe heterostructures with a thinner BP layer (Figure S5).

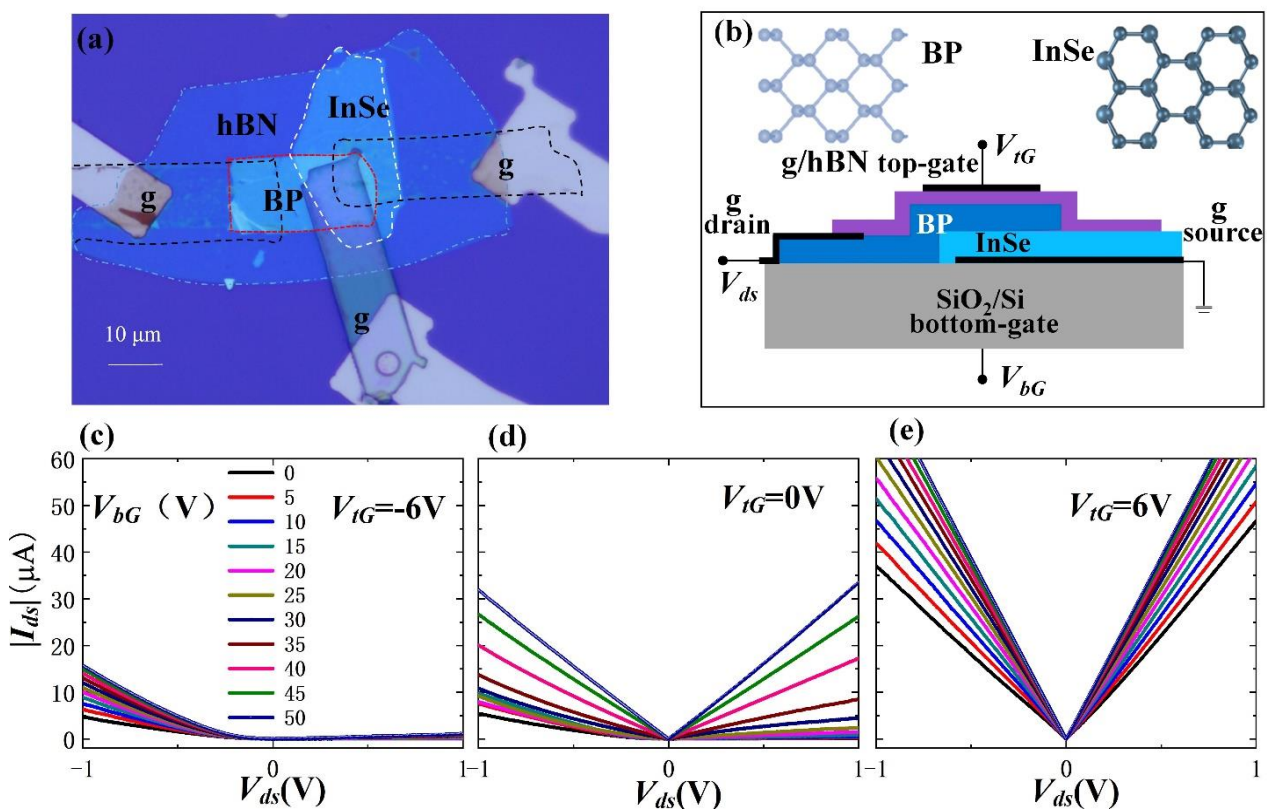

**Figure S7.** BP/InSe van der Waals heterostructure with a thick ($t_{BP} > 50$ nm) BP layer. Optical image (a) and schematic layout (b) of a BP/InSe FET on a $SiO_2$/Si substrate with graphene electrodes and a hBN encapsulating layer. The inset in part (b) shows the in-plane crystal lattice of BP and InSe. (c-e) Current-voltage $|I_{ds}|$-$V_{ds}$ curves at top gate voltages $V_{tG}$ = -8V (c), 0V (d), +8V (e) at room temperatures. Different curves in each panel correspond to bottom gate voltages $V_{bG}$ that vary from 0 to +50 V in steps of 5 V.

29